\documentclass{article}
\usepackage{spconf,amsmath,graphicx, url}
\usepackage{algorithmic}
\usepackage[ruled]{algorithm2e}

\title{Cascaded CNN-resBiLSTM-CTC: An End-to-End Acoustic Model For Speech Recognition}
\name{Xinpei Zhou, Jiwei Li, Xi Zhou\sthanks{\{zhouxinpei,lijiwei,zhouxi\}@cloudwalk.cn}}
\address{Cloudwalk Technology, Shanghai, P.R.China}
\begin{document}
%
\maketitle
\begin{abstract}
Automatic speech recognition (ASR) tasks are resolved by end-to-end deep learning models, which benefits us by less preparation of raw data, and easier transformation between languages. We propose a novel end-to-end deep learning model architecture namely cascaded CNN-resBiLSTM-CTC. In the proposed model, we add residual blocks in BiLSTM layers to extract sophisticated phoneme and semantic information together, and apply cascaded structure to pay more attention mining information of hard negative samples. By applying both simple Fast Fourier Transform (FFT) technique and n-gram language model (LM) rescoring method, we manage to achieve word error rate (WER) of 3.41\% on LibriSpeech test clean corpora. Furthermore, we propose a new batch-varied method to speed up the training process in length-varied tasks, which result in 25\% less training time.
\end{abstract}
\section{Introduction and Background}
\label{IB}
In the past decades, automatic speech recognition (ASR) tasks have been a major problem for scientists to tackle with. In the first decade of the new century, traditional MFCC-GMM-HMM models \cite{b17} had been the most popular pipeline to resolve ASR tasks. However, word error rate (WER) of this structure had remained above 20\%, which was highly unacceptable in real industry. With the booming development of deep learning in the recent 5 years, end-to-end deep learning model was first raised by Graves \cite{b1} in 2014, and significantly improved by Amodei \cite{b2} afterwards. 

This paper is inspired by Amodei \cite{b2}. Our main contributions in this paper to automatic speech recognition tasks include introducing the resBiLSTM(residual bidirectional long short-term memory) structure to replace the normal BiLSTM structure, a novel cascaded hard negative mining structure, and a more efficient training method. Furthermore, our model architecture can be easily transformed from English to any other languages (e.g. Mandarin, Spanish, France) as long as the corresponding vocabulary dictionary is provided. 

In this paper, we focus on designing an end-to-end deep learning acoustic model, it is more suitable and reasonable to use simple and computational friendly pre-processing and post-processing methods. Thus, Fast Fourier Transform (FFT) (pre-processing) and n-gram language model (LM) (post-processing) are applied in our structure, though Xiong \cite{b18} and Liu \cite{b8} showed RNN-LM outperforms n-gram LM. 

The structure of this paper is as follows. We start from the introduction on the recent related work on end-to-end ASR tasks in Section \ref{RW}. Section \ref{M} details the model architecture and improvements compared with the previous work. Then Section \ref{RA} comes up with the data preparation, experiment results, comparisons with other state-of-art end-to-end models, and relevant analysis. Section \ref{C} draws the conclusions and ideas on how to move forward in next steps.

\section{Related Work}
\label{RW}
In 2012, deep neural network (DNN) has taken advantages of increasing calculating powers, bigger training data, and better understanding on the models, has become more and more popular. DNN models proposed by Dahl \cite{b3} and Hinton \cite{b4} in 2012 significantly lower the error rate of SwitchBoard conversation data set for nearly $30\%$ compared with past MFCC-GMM-HMM models.

Based on those DNN models, end-to-end ASR models showed compelling results later in 2014. Connectionist Temporal classification (CTC) loss function proposed by Graves \cite{b5} had began to be applied by Graves \cite{b1}, Hannun \cite{b6}, and Maas \cite{b7} since 2014. CTC loss function gets rid of extracting phoneme information and making sentence alignment, which further improves the result compared with other DNN models without CTC. However, CTC is still unsatisfactory on homophone words judgment (e.g. brake and break). Thus, a subsequent LM rescoring mechanism was introduced later. Hannun \cite{b6} and Amodei \cite{b2} applied the tuned parameters between acoustic model, language model and sentence length to alleviate the CTC issues when decoding the sentence. Liu \cite{b8} also proposed alternative way called lattice rescoring by training a RNN-LM. 

As we intend to compare the results with existing models on the same data sets, we restrict the usage of raw training data. Since data augmentation has been proven beneficial to speech recognition tasks \cite{b9}, common data augmentation methods are applied to increase the versatility of data, such as speed perturbation ($0.9$x-$1.1$x) and noise disturbance. 

\section{Methodology}\label{M}
The whole system mainly consists of three parts: an extractor, an interpreter, and a corrector. The extractor extracts features from raw audio clips to spectrum. The interpreter interprets the spectrum into English words (Mandarin characters). The corrector corrects any grammar and spelling errors given by the interpreter to meaningful and complete sentences. 

Original raw audio clips are sliced into 20ms window (frame), with a sliding window of 10ms. For an audio clip with $x$ seconds, we have $N = \lfloor(1000x-20)/10\rfloor$ frames. For each frame, we use FFT to convert each frame into spectrum information with 161 features. $N*161$ features are formalized and sent to the interpreter, which will be illustrated in Section \ref{MA}. Other highlights of our model will be introduced in Section \ref{CS} to Section \ref{HT}. 

\begin{figure}[ht]
\centerline{\includegraphics[width=0.45\textwidth]{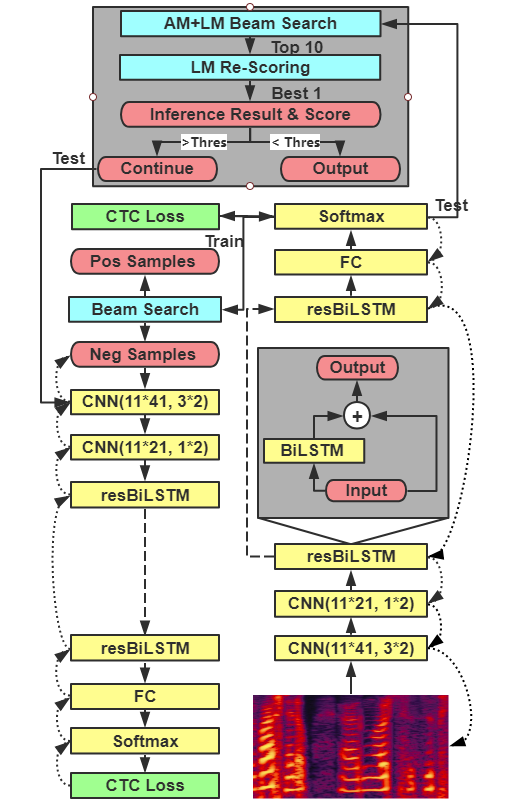}}
\caption{Illustration of detailed acoustic model architecture.}
\label{MS}
\end{figure}

\subsection{CNN-resBiLSTM-CTC acoustic model architecture}\label{MA}
Fig.~\ref{MS} show the detailed architecture of our CNN-resBiLSTM-CTC acoustic model. First several layers are normal CNN layers, which focus on extracting the phonemic information. Kernel size and stride size of each CNN layer are listed in each block in Fig.~\ref{MS}. Followed by CNN layers is a set of resBiLSTM layers. Our resBiLSTM layer consists of a shortcut that adds up the original information from the beginning of the layer and the information after BiLSTM. resBiLSTM layer is illustrated in the middle gray box in Fig.~\ref{MS}. Residual structure provides a fast flow path after the CNN layers, which allows the information extracted by CNN to be perceived by all LSTM layers, even the deepest layers. This structure allows the deeper LSTM layers to learn more sophisticated information by combining the phoneme and semantic information together, rather than learning the semantic information solely. Since without residual block, the phoneme information can be very hard to be passed to the deeper layers. The reason of using a BiLSTM structure is to get both forward and backward information within a sentence, as the word in a English sentence can be logically related either with previous or subsequent words. Then, one fully connected layer and one softmax layer are added after the resBiLSTM layers. Lastly, the whole network is trained via CTC loss function \cite{b3}. 

Suppose the length of the input sequence is $t$, and $q_t^{c_t}$ denotes the softmax probability of outputting label $c_t$, where $c_t\in\{a,b,c,...,z,\textit{blank}, \textit{space}, \textit{punctuations}\}$ at frame $t$. Denote $c=(c_1,...,c_t)$ as the CTC path of a sequence and $\theta(y)$ is the set of all possible CTC paths that can be mapped to sentence $y$. Therefore, the CTC loss $L_{ctc}$ is defined as in (1) and (2):
\begin{equation}
p_{ctc}(y|x) = \sum_{c\in\theta(y)}p_{ctc}(c|x) = \sum_{c\in\theta(y)}\prod_{t=1}^Tq_{t}^{c_t}\label{eq1}
\end{equation}
\begin{equation}
L_{ctc} = -\textit{ln}(p_{ctc}(y|x)).\label{eq2}
\end{equation}

\subsection{Cascaded structure and hard negative mining}\label{CS}
Cascaded CNN model has been proved successfully on computer vision tasks. Li \cite{b10} applied the idea on face detection and Sun \cite{b11} made progress on facial point detection. We hold the idea that cascaded models can also be transfered to CNN-RNN combined models.

After training on the original data set until converge, we make inference on the training set again using the converged model. Then, we inference the audio clips in testing set with the converged model via beam-search and select the samples for the second-stage via cascaded structure. Pseudo-code below shows the detailed methods we use. We apply almost the same procedure on training set again to select the wrong samples for second-stage, except we don't use language model here.

\begin{algorithm}
\caption{Beam search and cascaded structure}
\For{audio $x_i$ with $T_i$ time stamps}
{
    \textit{top N candidates set S = } \{\}  \\
    \For{$t in 1,2,3,...,T_i$}
    {
        Calculate, select top N $q_t^{c_t}$ given $s_k$ in $S$ \\
        From k*N probabilities, select top N collections $s_1, ..., s_N$ and update S\\
    }
    \textit{select top 10 probability collections from C} \\
    \eIf{$p_{ctc}(s_1|x)/len(s_1) > thres$}{
        \textit{goto cascaded model}
    }
    {
        \textit{combine with LM, select the best one}
    }
}
\end{algorithm}

We figure out $43\%$ utterances are not being recognized correctly using trained acoustic model solely. In order to find the specific differences between wrong samples and all samples, we compare them on the following attributes in Table \ref{tab4}: average word length, average speaking speed, and each character's appearance rate.

Among all characters, $k$ is the most biased one ($3.41\%$ difference), which still distributes insignificantly different. Similarly, the speaking speed differs insignificantly, implying unnecessity of altering stride size in CNN layers to capture different spectrum and phoneme information in the second-stage. However, the average word length increases by $11.0\%$ on wrong samples. As the word length increases, highly semantic correlated words are even more distant. Belinkov \cite{b12} shows deeper resBiLSTM layers contributes more on capturing semantic information, we increases the number of layers from 7 to 13 in the second-stage cascaded model that focus on those wrong samples. Since training samples decrease for 43\% in second-stage, we decrease the model size accordingly to accommodate this change and modify the hidden node size to 512. Since the cascaded structure focuses on solving semantic issues of longer sentences, rather than phoneme problems. Also, considered that we use the same 2-layer CNN structure in both stages, we use the trained CNN weights from the first-stage model directly and start training second-stage model from there.\\

\begin{table}[ht]
\caption{Statistics of wrong and all samples.}
\begin{center}
\begin{tabular}{|c|c|c|c|}
\hline
\textbf{Attributes} & \textbf{\textit{Wrong}}& \textbf{\textit{All}} & \textbf{\textit{Diff.}}\\
\hline
Avg. length (words/sen) & 13.64& 12.30 & \textbf{+11.0\%}\\
\hline
Avg. speed (chars/sec) & 14.55& 14.39 & +1.11\%\\
\hline
Appearance rate of e (\%) & 10.24& 10.31& -0.64\%\\
\hline
Appearance rate of l (\%) & 3.32& 3.30& +0.69\%\\
\hline
Appearance rate of k (\%) & 0.65& 0.63& +3.41\%\\
\hline
\end{tabular}
\label{tab4}
\end{center}
\end{table}

\subsection{Varied batch size and length sorting}\label{VBS}
Since the lengths of audio clips vary, training with mini-batches would inevitably introduce zero padding to make sentence alignment. However, padded zeroes are useless in training and could cause gradient vanishing problems while training LSTM structure is very subtle. To wipe off this extra issue, we want to minimize the number of padded zeroes. First, we could sort the training audio clips in ascending order and each mini-batch takes sorted consecutive $k$ samples. Therefore, the difference between shortest and longest time-stamp would be smallest in each mini-batch, which result in least padded zeroes. Second, Amodei \cite{b2} use the same batch size throughout the experiment, which would cause waste of memory usage. In LibriSpeech training set, the longest time-stamp could be 10 times longer than the smallest time-stamp. In order to prevent out-of-memory issue, we need to stick to batch size that suits the last mini-batch if we set the batch size consistently. However, this would cause nearly 90\% of the memory wasted when training the first several batches. We come up with a solution by adjusting the batch size accordingly with its longest batch time-stamp, by computing the ratio of the longest length in each batch. When setting biggest batch size to 5 times bigger than the smallest batch size, we increase the utility of GPU memory usage for up to $25\%$. In the mean time, the training time for each epoch decreases 24.9\% from 24500 seconds to 18400 seconds.

\subsection{Decoding with external n-gram language model}\label{LM}
The language model training corpora is available\footnote{\url{http://openslr.org/11/}} online, which contains 14500 public domain books with more than 41 million sentences. We eliminate sentences that contains any words appear less 100 times in the corpora, and keep 39 million sentences. We take advantage of KenLM \cite{b13} to build 3-gram, and 5-gram language models respectively. Language models are used to re-score the decoded sentences. 

\subsection{Hyper-parameters tuning}\label{HT}
When we do inference, the result given by acoustic model (AM) is paired with the n-gram LM mentioned in Section \ref{LM}. We use the beam search \cite{b14} to derive the transcription t with minimum $S(t)$, where
\begin{equation}
S(t) = \log(p_{AM}(t|x)) + \alpha \log(p_{LM}(t))\label{eq3}
\end{equation}
, and $\alpha$ is the hyper-parameter to be tuned. We randomly select 50 numbers in $[0,5]$ of $\alpha$. If we define WER $=f(\alpha)$, This is a convex function, and WER reaches global minimum at $\alpha=2.0$. However, $\alpha$ reaches different global minimum for different AM or LM models. One has to re-tune the hyper-parameter after changing either AM or LM.

\section{Results and Analysis}\label{RA}
All experiments are carried out on a 8-GPU Nvidia 1080Ti server on LibriSpeech data sets, containing 960-hour reading audio clips with 281241 utterances. Training process takes 4.5 days without cascaded structure and extra 1.5 days with cascaded structure. We exclude audio clips longer than 21 seconds to speed up training process, which causes only $0.019\%$ data loss. Minimal batch size is set at 48 on 7-layer resBiLSTM structure with 1024 hidden nodes each. All models are training for 20 epochs and early termination is applied. Learning rates are set to $5*10^{-4}, 5*10^{-5}, 5*10^{-6}$ in the first 10 epochs, next 5 epochs and final 5 epochs respectively. Momentum $=0.99$ and stochastic gradient descent with Adam are fixed. When testing, beam search width is set to be 300. Different LM size are applied in the experiment, we report no-gram, 3-gram, and 5-gram results respectively. We conduct experiments with and without the second-stage cascaded model and residual block to justify their improvements. Hard samples with $p_{AM} > 0.5$ are selected into second-stage cascaded model. 840/2620, 1031/2939, 860/2703, and 1002/2864 hard samples are selected from test-clean, test-other, dev-clean, and dev-other data sets respectively.

\begin{table}[ht]
\caption{WERs on different structure settings and LM size.}
\begin{center}
\begin{tabular}{|c|c|c|c|c|c|}
\hline
AM& LM& \multicolumn{2}{|c|}{\textbf{Test}}& \multicolumn{2}{|c|}{\textbf{Dev}} \\
\cline{3-6} 
strcuture & size& \textit{clean}& \textit{other}& \textit{clean}& \textit{other}\\
\hline
7(Baseline, BL)&5& 3.85&12.48&3.77&11.62\\
\hline
7&3&4.10&12.61&3.89&11.77\\
\hline
7&0&5.64&16.79&5.47&16.43 \\
\hline
13&5&3.97&12.68&3.83&11.96 \\
\hline
BL+13-cascade&5&3.81&12.27&3.75&11.48\\
\hline
BL+residual&5&3.45&11.01&3.43&10.72\\
\hline
BL+both&5& \textbf{3.41}& \textbf{10.65}& \textbf{3.39}& \textbf{10.45} \\
\hline

\end{tabular}
\label{tab1}
\end{center}
\end{table}

Our baseline setting has 7-layer BiLSTM and 5-gram LM. As language model size increases, WER drops accordingly. We add a cascaded structure with 13-layer BiLSTM with 512 hidden nodes each to the baseline settings, WER drops from 3.85\% to 3.81\%. Enhancement on adding residual block is significant, WER drops from 3.85\% to 3.45\%. If we combine both cascaded structure and residual block together, WER further drops to 3.41\%. To prove that the better performance of second-level cascaded structure is not related to the deeper LSTM layers, we conduct an experiment using all training samples with 13-layer resBiLSTM structure only. Result is even worse than the cascaded structure (3.97\% vs 3.81\%). 

\begin{table}[h]
\caption{Comparison on all/hard samples of test-clean data.}
\begin{center}
\begin{tabular}{|c|c|c|c|c|}
\hline
AM&LM&\textit{All}& \textit{Hard}& \textit{Wrong}\\
structure&size&\textit{(Errors)}& \textit{(Errors)}& \textit{hard/all}\\
\hline
7(BL)&5& 3.85(2024)&6.29(1316)&65.0\%\\
\hline
13&5&3.97(2089)&6.38(1335)&63.9\%\\
\hline
+cascade&5&3.81(2001)&6.18(1293)&64.6\%\\
\hline
+residual&5&3.45(1815)&5.64(1179)&65.0\%\\
\hline
+both&5&3.41(1792)&5.53(1156)&64.5\%\\
\hline

\end{tabular}
\label{tab10}
\end{center}
\end{table}

\begin{table}[h]
\caption{Results on different end-to-end model structures.}
\begin{center}
\begin{tabular}{|c|c|c|c|c|}
\hline
\textbf{Model}&\multicolumn{2}{|c|}{\textbf{Test}}& \multicolumn{2}{|c|}{\textbf{Dev}} \\
\cline{2-5} 
(\# parameters) & \textit{clean}& \textit{other}& \textit{clean}& \textit{other}\\
\hline
Baidu DS2 (109M)\cite{b2}& 5.15& 12.73& -& - \\
\hline
ESPnet (134M)\cite{b15} & 4.0& 12.8& 3.9& 11.8 \\
\hline
I-Attention \cite{b16} & 3.82& 12.76& 3.54& 11.52 \\
\hline
Residual (119M)& \textbf{3.45}& \textbf{11.01}& \textbf{3.43}& \textbf{10.72} \\
\hline
Both (174M) & \textbf{3.41}& \textbf{10.65}& \textbf{3.39}& \textbf{10.45} \\
\hline
\end{tabular}
\label{tab2}
\end{center}
\end{table}

To investigate the effect of cascaded structure, we further examine the performance of different models on the hard samples solely as cascaded models do not contribute to those easy samples. Take test-clean data set as an example,we compare the performances in Table \ref{tab10}. Average length for hard samples is 24.91 words/sentence while the average is only 20.07 for all samples. Cascaded structure decreases WERs on wrong samples for 0.11\% in both cases (6.29\%/6.18\%, 5.64\%/5.53\%). Residual block performs better on both data sets compared with cascaded structure. However, the relative increment on hard samples is slightly worse than cascaded structure (65.0\% vs 64.6\%), which proves cascaded models do have positive impact on hard samples. Another notable finding is that 13-layer single-stage structure performs better on hard samples out of all samples relatively (63.9\% vs 65.0\%), however, it performs worse absolutely compared to the baseline.

We compare our results with other end-to-end models in Table \ref{tab2}. We get better result with roughly the same amount of parameters using residual block. By adding cascaded structure, we further increase our performance.

\section{Conclusion}\label{C}
This paper enhances the performances based on the work proposed by Amodei \cite{b2}. By adding residual network structure in LSTM layers to extract more abstract and sophisticated phoneme and semantic information, adding cascaded model structure to categorize hard samples better, and proposing new methods to speed up training process, we reach state-of-art result on LibriSpeech corpora.

Based on the achievements we have gained, we need to focus on training an end-to-end AM+LM model using the same corpora as current training structure actually causes information loss when combining independent AM and LM parts. Another potential enhancement might be ameliorating RNN structure, such as adding frame skipping structure to capture more semantic information. Also, our team is planning to apply the same structure on self-collected Chinese data soon. 


\bibliographystyle{IEEEbib}
\bibliography{refs}

\end{document}